\documentclass[12pt,preprint]{aastex}

\slugcomment{To appear in The Astrophysical Journal}

\shorttitle{Collimated jet towards IRAS16547-4247}
\shortauthors{Rodr\'\i guez et al.}

\begin{document}

\title{High Angular Resolution Observations of the 
Collimated Jet Source Associated with a Massive Protostar
in IRAS~16547-4247}

\author{Luis F. Rodr\'\i guez}
\affil{Centro de Radioastronom\'\i a y Astrof\'\i sica, UNAM,
Apdo. Postal 3-72, Morelia, Michoac\'an, 58089 M\'exico}
\email{l.rodriguez@astrosmo.unam.mx}

\author{Guido Garay}
\affil{Departamento de Astronom\'\i a,
Universidad de Chile, Casilla 36-D, Santiago, Chile}
\email{guido@das.uchile.cl}

\author{Kate J. Brooks}
\affil{Australia Telescope National Facility, P.O. Box 76, Epping NSW 1710
Australia}
\email{Kate.Brooks@atnf.csiro.au}

\and

\author{Diego Mardones}
\affil{Departamento de Astronom\'\i a,
Universidad de Chile, Casilla 36-D, Santiago, Chile}
\email{mardones@das.uchile.cl}

\begin{abstract}

A triple radio source recently detected in association with the luminous infrared
source IRAS~16547-4247 has been studied with high angular resolution and
high sensitivity with the Very Large Array at 3.6 and 2 cm.
Our observations confirm the interpretation that the central object is
a thermal radio jet, while the two outer lobes are most probably
heavily obscured HH objects.
The thermal radio jet is resolved angularly for the first time and
found to align closely with the outer lobes. 
The opening angle
of the thermal jet is estimated to be $\sim 25^\circ$, confirming that 
collimated outflows can also be present in massive protostars.
The proper motions of the outer lobes should be measurable over timescales
of a few years.
Several fainter sources detected in the region are most probably associated
with other stars in a young cluster.
\end{abstract}


\section{Introduction}

Perhaps the major question related with galactic star formation is 
whether or not the successful model of low-mass star formation,
based on accretion via a circumstellar disk and collimated
outflow in the form of jets (Shu, Adams, \& Lizano 1987), can be
extended to the case of high-mass protostars.
A handful of B-type young stars have been found to be associated
with collimated outflows and even possibly circumstellar disks
(see Garay \& Lizano 1999). The source studied here,
IRAS~16547-4247, is possibly the best 
case of a highly collimated outflow associated with an O-type protostar.

IRAS~16547-4247 is a region of star formation located at a distance of 2.9 
kpc. It has a bolometric luminosity of 6.2$\times$10$^4$ $L_\odot$, 
equivalent to that of a single O8 zero-age main-sequence star, although
most probably a cluster is present and 
the most massive star is of slightly lower luminosity. Garay et al. 
(2003) recently detected an embedded triple radio continuum source
associated with the IRAS source.
The three radio components are aligned in a northwest-southeast direction, 
with the outer lobes symmetrically separated from the central source by 
an angular distance of $\sim 10{''}$, equivalent to
a physical separation in the plane of the sky of $\sim$0.14 pc. 
The positive spectral index of the central source
is consistent with that expected for a radio thermal (free-free) jet
(e. g. Anglada 1996; Rodr\'\i guez 1997),
while the spectral index of the lobes suggests non thermal emission.
The triple system is centered on the position of the IRAS source and is 
also coincident within error with a 1.2 mm dust continuum and
molecular line emission core whose
mass is of order 10$^3$ $M_\odot$ (Garay et al. 2003).
Brooks et al. (2003) reported a chain of $H_2$ 2.12 $\mu$m emission 
knots that trace a collimated flow extending over 1.5 pc
that emanates from close to the central component of the triple radio
source and has a position angle very similar to that defined by the
outer lobes of the triple radio source. Most likely this extended
component traces gas ejected in the past by the central component of
the triple source. 

In this paper we present sensitive, high angular resolution Very Large Array 
observations and new ATCA radio continuum observations that provide new 
information on the characteristics of the radio triple source in IRAS~16547-4247.
 
\section{Observations}

The observations were made using the Very Large Array (VLA) of the National
Radio Astronomy Observatory (NRAO)\footnote{NRAO is a facility of the
National Science Foundation operated under cooperative agreement by
Associated Universities, Inc.} and the Australia Telescope Compact Array (ATCA)
\footnote{The Australia Telescope Compact Array is funded by the Commonwealth of
Australia for operation as a National Facility managed by CSIRO.} in Australia.

\subsection{Very Large Array}

The VLA radio continuum observations were carried out in 2003 September 25 
and 30, at the frequencies of 8.46 and 14.9 GHz. The array was in the BnA 
configuration and an effective bandwidth of 100~MHz with two circular
polarizations was employed. The absolute amplitude calibrator was 1331+305 
(with adopted flux densities of 5.21 and 3.46 Jy, at 8.46 and 14.9 GHz, 
respectively). 
At 14.9 GHz the 1331+305 data were used in conjunction with a source model 
provided by NRAO. The phase calibrator was 1626$-$298, with the bootstrapped 
flux densities given in Table 1. The phase 
center of the array was $\alpha=16^h58^m17{\rlap.}{^s}202$ and 
$\delta=-42^\circ52'09{\rlap.}{''}59$ (J2000.0). The 
data were edited and calibrated using 
the software package Astronomical Image Processing System (AIPS) of NRAO. 
To correct for amplitude and phase noise induced by the low elevation of 
the source, the data were self-calibrated.  No significant variations were
found between the two epochs of observations (separated by only five days) 
and the final analysis was made concatenating all data. Cleaned maps were 
obtained using the task IMAGR of AIPS.
The synthesized (FWHM) beams 
are $1.13\arcsec\times0.61\arcsec$ and $0.95\arcsec\times0.42\arcsec$ at the
frequencies of 8.46 and 14.9 GHz, respectively. The noise level achieved in the
images are 30 and 87 $\mu$Jy beam$^{-1}$ at 8.46 and 14.9 GHz, respectively.

\subsection{ATCA}

The ATCA radio continuum observations were made in 2003, February 28,
using the 6A configuration, which utilizes all six antennas and covers 
east-west baselines from 300 m to 5.9 km. Observations were made at the 
frequencies of 4.80 and 8.64 GHz, each with a bandwidth of 128 MHz. The phase 
center of the array was $\alpha=16^h58^m16{\rlap.}{^s}847$ and 
$\delta=-42^\circ51'37{\rlap.}{''}23$ (J2000.0). The total integration time 
at each frequency was 285 minutes, obtained from 15-minute scans taken over 
a range of hour angles to provide good (u,v) plane coverage. The calibrator 
1616-52 was observed before and after every on-source scan in order to correct 
the amplitude and phase of the interferometer data for atmospheric and 
instrumental effects. The same calibrator was also used for the bandpass 
correction. The flux density was calibrated by observing PKS 1934-638 (3C84) 
for which values of 5.83 and 2.84 Jy at 4.8 and 8.6 GHz, respectively, were 
adopted. Standard calibration and data reduction were performed using MIRIAD 
(Sault, Teuben, \& Wright 1995).
Maps were made by Fourier transformation of the uniformly weighted
interferometer data using the AIPS task MX.  The synthesized (FWHM) beams 
are $4.08\arcsec\times1.49\arcsec$ and $2.32\arcsec\times0.80\arcsec$ at the
frequencies of 4.80 and 8.64 GHz, respectively. The noise level achieved in the
images are 70 and 72 $\mu$Jy beam$^{-1}$ at 4.80 and 8.64 GHz, respectively.

\section{Results and Discussion}

Figures 1 and 2 present contour maps of the emission observed, respectively, 
with ATCA (4.80 and 8.64 GHz) and with the VLA (8.46 and 14.9 GHz). These images 
show that there are three dominant sources that Garay et al. (2003) identified as 
the central source and the two (north and south) outer lobes. 
The high sensitivity and angular resolution of the 8.46 GHz VLA image allowed us 
to resolve the spatial structure of these features and to detect several other 
components, which will be discussed in what follows.
Table 2 presents the positions, flux densities and deconvolved
angular dimensions of the sources associated with the triple radio continuum 
system (jet and two lobes). The observed parameters were determined from a 
linearized least-squares fit to a Gaussian ellipsoid function using the task 
IMFIT of AIPS.

We note that the 8.46 GHz flux densities measured for the three main sources
with the VLA in 2003 September are similar (within a factor of 1.2)
to the 8.64 GHz flux densities measured with ATCA in 2003 February and larger by
factors of 1.5 to 2 with respect to the 8.64 GHz flux densities reported by Garay
et al. (2003) from observations taken with ATCA in 2000 May. Although we cannot
rule out a calibration problem, we believe that this difference could be due to
the well known fact that both thermal jets (Rodr\'\i guez et al. 2001;
Galv\'an-Madrid, Avila, \& Rodr\'\i guez 2004) and HH knots (Rodr\'\i guez et
al. 2000) can show significant flux density variations on scales of years or
even months. 

\subsection{The central source}

From the ATCA observations we measured that the central source has flux densities 
of $7.3\pm0.4$ mJy at 4.8 GHz and $8.2\pm0.2$ mJy at 8.64 GHz. From the VLA 
observations we measured flux densities of $8.7\pm0.1$ mJy at 8.46 GHz and 
$10.5\pm0.3$ mJy at 14.9 GHz. Assuming that the flux density has a power law 
dependence with 
frequency of the form $S_\nu \propto \nu^\alpha$, we derive that the central 
source has a spectral index of 0.33$\pm$0.05 between 8.46 and 14.9 GHz, similar 
to the value of 0.49 derived by Garay et al. (2003) and consistent with free-free 
emission from a collimated jet. 

The central source is clearly resolved angularly with the VLA (see Figure 2).
We determined deconvolved major and minor axis of $1.02\arcsec\times0.22\arcsec$, 
P.A. of 163$^{\circ}$, at 8.46 GHz, and $0.74\arcsec\times0.22\arcsec$, P.A. 
167$^{\circ}$, at 14.9 GHz. Assuming that the angular size of the major axis of 
the central source has a power law dependence with frequency
of the form $\theta_\nu \propto \nu^\beta$, we find that this object has a 
power law angular size index of -0.56$\pm$0.05, consistent with the
values expected for a thermal jet source (Reynolds 1986), whose angular
dimensions diminish with increasing frequency.
We assume that the
opening angle of the thermal jet is the angle subtended  
by the deconvolved minor axis at a distance of one half the
deconvolved major axis (Eisl\"offel et al. 2000), namely that the
opening angle is two times the arctan of the ratio of the deconvolved minor axis
over the deconvolved major axis. 
Using the 8.46 GHz deconvolved angular dimensions given in Table 2,
we then estimate the opening angle of the thermal
jet to be $\sim$25$^\circ$, indicating significant collimation
in this massive protostar.

\subsection{The north lobe}

The 8.46 GHz VLA image shows that the north lobe has a string-like
structure, that almost connects the north tip of the lobe with the central source. 
The structure is composed of at least three components, labeled N-1, N-2 and N-3, 
which are also seen in the 8.64 GHz ATCA map. These three components 
are along the outflow axis and we will consider them to be knots in the jets. 
At 14.9 GHz, emission from the north lobe 
is detected only from the brighter 8.46 GHz knot (component N-1). 

Due to the different angular resolution at the different frequencies and the 
blending of emission from the different components in the northern lobe it is not 
straightforward to derive the spectral index of their emission. To partially 
avoid this problem we made tapered maps of the emission at the higher frequencies 
observed with ATCA and with the VLA. The taper value was chosen to match the beam 
at the lower frequency. Using the ATCA data we measure at the 
peak position of the north lobe peak flux densities of 2.67 mJy beam$^{-1}$ at 4.80 GHz 
(HPBW of $4.08\arcsec\times1.49\arcsec$) and 2.45 mJy beam$^{-1}$ at 8.64 GHz (HPBW of 
$4.05\arcsec\times1.48\arcsec$), implying an spectral index of $-0.15\pm0.09$. 
From the VLA observations, we find that the N-1 component has an
spectral index of 0.17$\pm$0.39 between 8.46 and 14.9 GHz. These results, 
indicating a flat spectrum, suggest that the radio continuum emission at the 
peak position  of the northern lobe corresponds to optically thin thermal emission. 
The spectral index of the integrated emission from the whole northern lobe is 
$-0.32\pm0.29$.

\subsection{The south lobe}

The 8.46 GHz VLA image shows that the south lobe is composed of a bright component, 
labeled S-1, and a weak component, labeled S-2. No emission is detected towards 
this lobe connecting the bright component with the central source.
The triple source in IRAS~16547-4247 is similar to the
triple source in Serpens (Curiel et al. 1993), in the sense that both
show one side with many knots connecting a lobe with the central source,
while there is no detectable emission between the central source and the other
lobe.

Using the ATCA data set with similar beams, we determined a spectral index
between 4.8 and 8.64 GHz of $-0.36\pm0.09$ for the peak flux density and of 
$-0.65\pm0.14$ for the integrated emission from S-1.
Using the VLA data, we measure flux densities of $2.8\pm0.1$ mJy at 8.46 GHz 
and $2.0\pm0.1$ mJy at 15 GHz, implying a spectral index of -0.59$\pm$0.15.
These results suggest that the emission from the southern lobe has a non thermal 
origin. A detailed discussion 
on the presence of thermal and non thermal components in this type of objects is
given in Garay et al. (1996).
The duality in emission mechanisms is expected in shock waves where a small 
fraction of the electrons are accelerated to relativistic velocities, giving 
rise to nonthermal emission, while most of the electrons produce thermal 
free-free emission (Crusius-W\"atzel 1990; Henriksen, Ptuskin, \& Mirabel 1991).
It is interesting to note that the outflows where a non thermal component has been found
(Serpens: Rodr\'\i guez et al. 1989, HH~80-81:
Mart\'\i, Rodr\'\i guez, \& Reipurth 1993, Cep A: Garay et al. 1996, 
W3(OH): Wilner, Reid, \& Menten 1999, IRAS~16547-4247: this paper)
are all associated with massive young stellar objects.


\subsection{Alignment and proper motions}

For each of the three main sources, the position angles of the major axis 
determined with the VLA at 8.46 and 14.9 GHz are consistent within error. 
The position angle of the major axis of the central thermal jet is, from the 
more accurate 8.46 GHz data, 163$^\circ \pm$1$^\circ$. This position angle 
agrees extremely
well with the position angle of 163$^\circ \pm$1$^\circ$ determined from the 
8.46 GHz positions given in Table 2 for the N-1 and S-1 components. 
However, as first noted by Garay et al. (2003), there is a slight misalignment 
between the lobes and the central source (see Figure 2), in the sense that 
the central source is displaced by about 0$\rlap.{''}$5 to the northeast of 
the line that joins the lobes. Garay et al. (2003) propose that this 
displacement could be explained in terms of a jet that is ejected from a source
with an orbital motion (Masciadri \& Raga 2002).
Similar small misalignments have been found in other thermal jet sources:
L1551~IRS5 (see Figure 1 of Rodr\'\i guez et al. 2003),
HH~1-2 (see Figure 1 of Rodr\'\i guez et al. 2000),
and HH~80-81 (see Figure 1 of Mart\'\i, Rodr\'\i guez, \& Reipurth 1993 and
Figures 2 and 3 of Mart\'\i, Rodr\'\i guez, \& Reipurth 1998).
Since there are only a handful of triple radio sources where both lobes and the 
central source are simultaneously detected, this misalignment appears to be 
rather common among these sources and deserves further investigation. 

Given the high signal to noise ratio of the image,
the relative positions of the sources can be determined with great
accuracy from the images. From the 8.46 GHz image we get an angular separation
of $20\rlap.{''}38 \pm 0\rlap.{''}02$ between the N-1 and S-1 components,
whereas at 2 cm this angular separation is found to be $20\rlap.{''}40 \pm
0\rlap.{''}11$. Assuming that the knots move at 500 km s$^{-1}$ in the
plane of the sky, we expect their separation to increase by $0\rlap.{''}2$ in
a three year period and we should be able to determine proper motions at a
10-$\sigma$ accuracy by obtaining a new 8.46 GHz image in 2006.
It is very important to measure the proper motions of these knots, since the
detection of large values will corroborate the interpretation that a massive
protostar is producing the outflow. Jets from massive protostars are
characterized by larger velocities 
(500-1000 km s$^{-1}$; Mart\'\i, Rodr\'\i guez, \& Reipurth 
1995) than jets from
low mass protostars (where velocities below 300 km s$^{-1}$ are the rule;
e. g. Reipurth et al. 2000, Bally et al. 2002).

\subsection{Other sources in the field}

In the 14.9 GHz VLA image (see Figure 2) we only detect the central source
and the brighter components of the lobes (N-1 and S-1) above the 5-$\sigma$ level. 
The 8.46 GHz VLA image is several times more sensitive and besides the central 
source and the two outer lobes, three other components, labeled A, B and C, are 
detected.  They are clearly displaced from the main axis of the outflow.
Their positions and flux densities are given in Table 3.

Source A, located $\sim 4\arcsec$ northeast of N-1, seems to be part of a complex 
structure associated with the northern lobe. Emission from this source was 
detected with ATCA at 4.8 and 8.64 GHz, with flux densities of $1.7\pm0.3$
and $1.2\pm0.4$ mJy, respectively. The flux density measured with the VLA 
at 8.46 GHz is about a third of that measured with ATCA at 8.64 GHz, 
suggesting that most of the emission is resolved out with the high angular 
resolution VLA observations.

Source B was detected with the VLA at 8.46 GHz (September, 2003), with a flux 
density of 0.67 mJy, but it was below the detection limits at 8.64 GHz when 
observed with ATCA (February, 2003). Garay et al. (2003) detected source B at 
8.64 GHz with a flux density of 0.9 mJy during May, 2000.
These observations indicate that source B is time variable. In addition, 
we detected emission from source B at 4.8 GHz with ATCA with a flux density
of 0.44 mJy.

Source C was detected with ATCA at 4.8 and 8.64 GHz, with flux densities of 
1.08$\pm$0.05 and 0.74$\pm$0.06 mJy, respectively. It was clearly detected with 
the VLA at 8.46 GHz, with a flux density of 0.90 mJy, and marginally detected at 
14.9 GHz, with a flux density of $0.64\pm0.20$ mJy. Source C is not present in 
the 8.64 GHz image of Garay et al. (2003) indicating it is time variable.
The spectral index derived from the 4.8 and 8.64 GHz ATCA flux densities 
is $-0.6\pm$0.2, in good agreement with the less reliable spectral index 
between 8.46 and 14.9 GHz derived from the VLA data,
suggestive of a non thermal spectrum. In addition, 
source C was found to be unresolved in all the observed frequencies. 
We propose that this source could be associated with a young, 
low-mass star with gyrosynchrotron emission (see G\'omez, Rodr\'\i guez, \& 
Garay 2002).

\acknowledgments

LFR is grateful to 
CONACyT, M\'exico and DGAPA, UNAM for their support. DM and GG acknowledge 
support from the Chilean {Centro de Astrof\'\i sica} FONDAP 15010003.

\clearpage

\begin{figure}
\plottwo{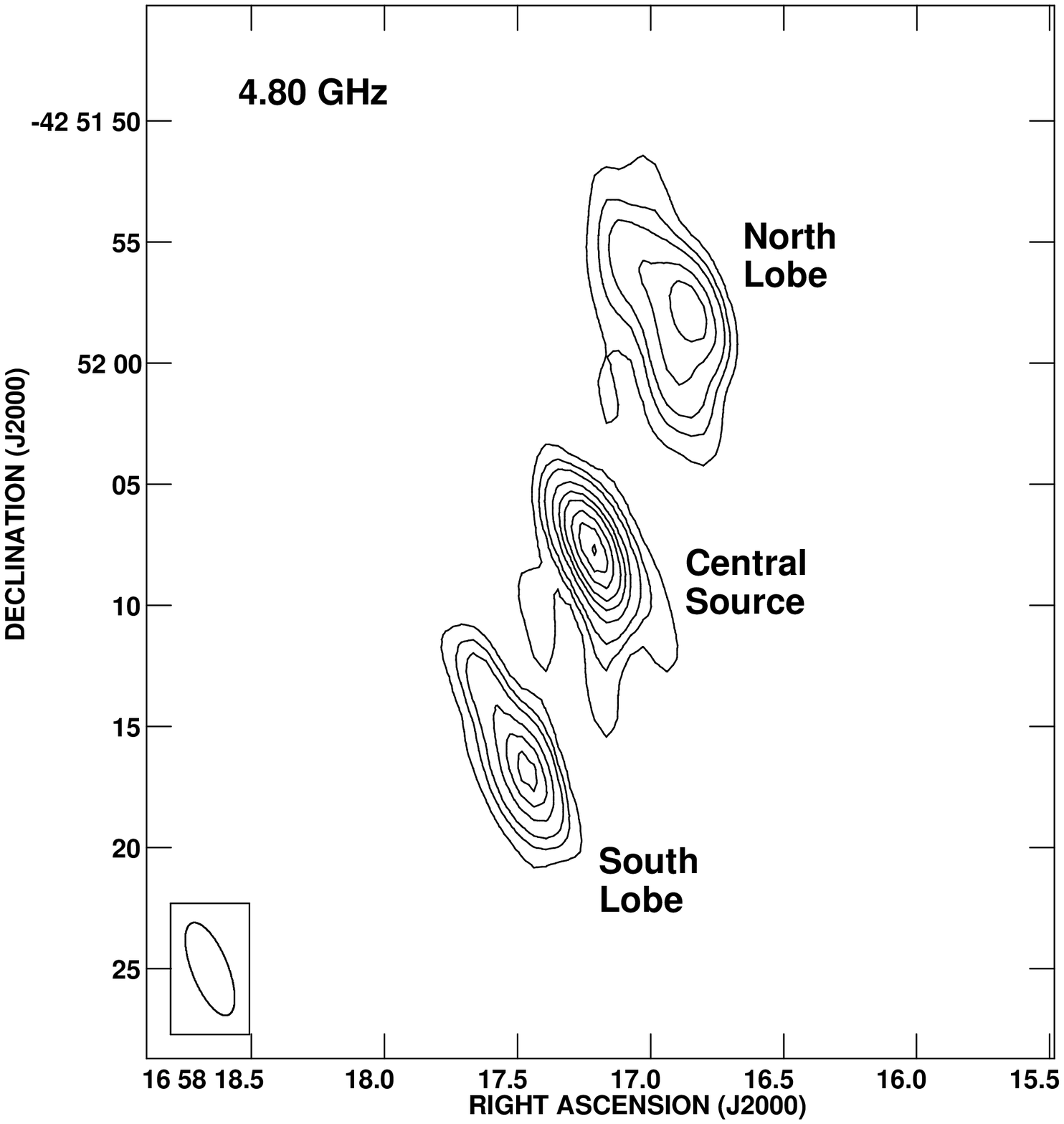}{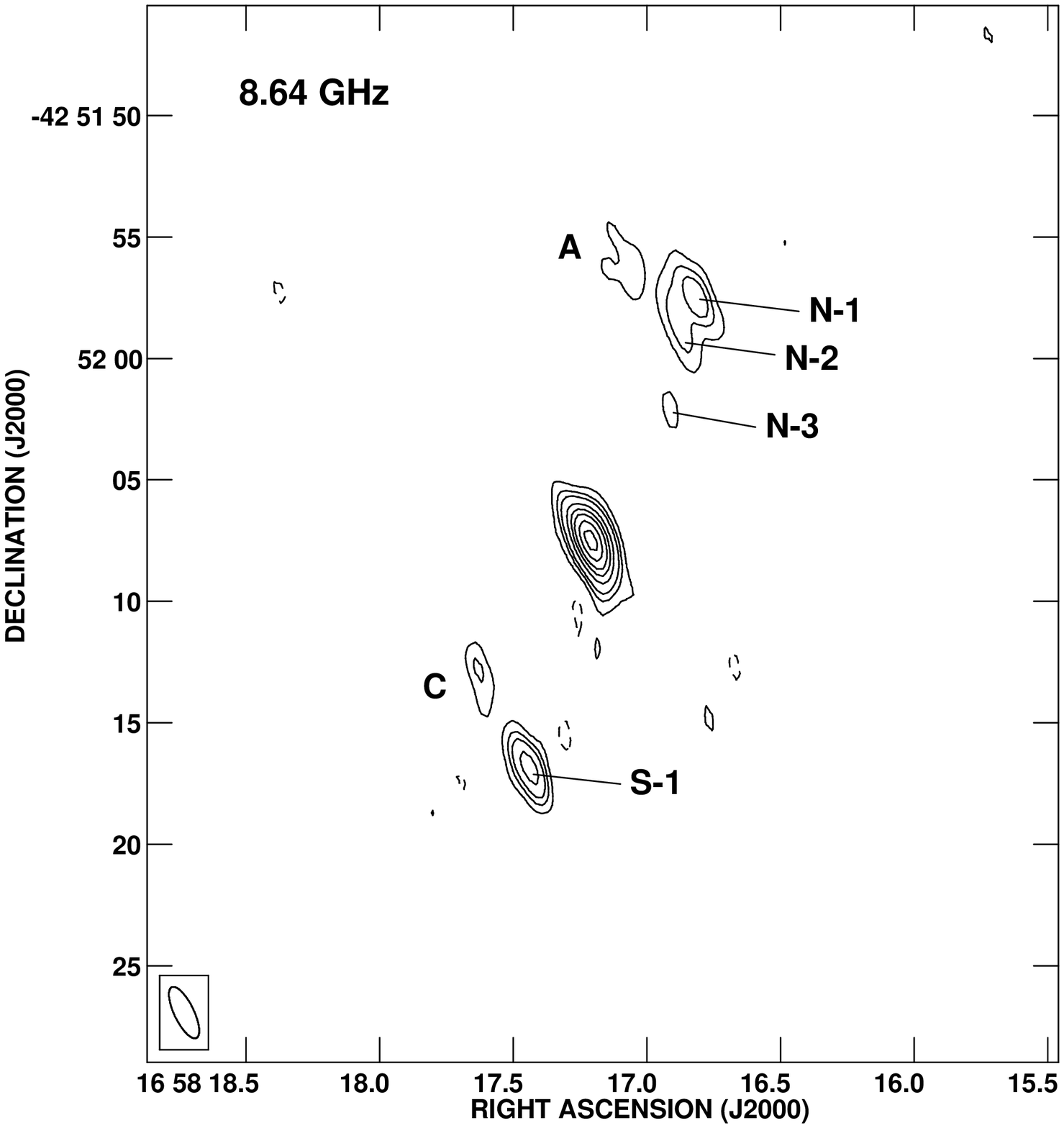}
\caption{ATCA images at 4.80 (left) and 8.64 GHz (right) towards 
IRAS~16547-4247.
Contours are -1, 1, 2, 3, 5, 7, 9, 12, 15, and 18 times
0.3 mJy beam$^{-1}$. The half power contour of the synthesized beams
($4\rlap.{''}08 \times 1\rlap.{''}49$; PA = $22^\circ$ for the 
4.8 GHz image and $2\rlap.{''}32 \times 0\rlap.{''}80$; PA = $26^\circ$ 
for the 8.6 GHz image) are shown in the bottom left corner of each panel.
\label{fig1}}
\end{figure}

\clearpage

\begin{figure}
\plottwo{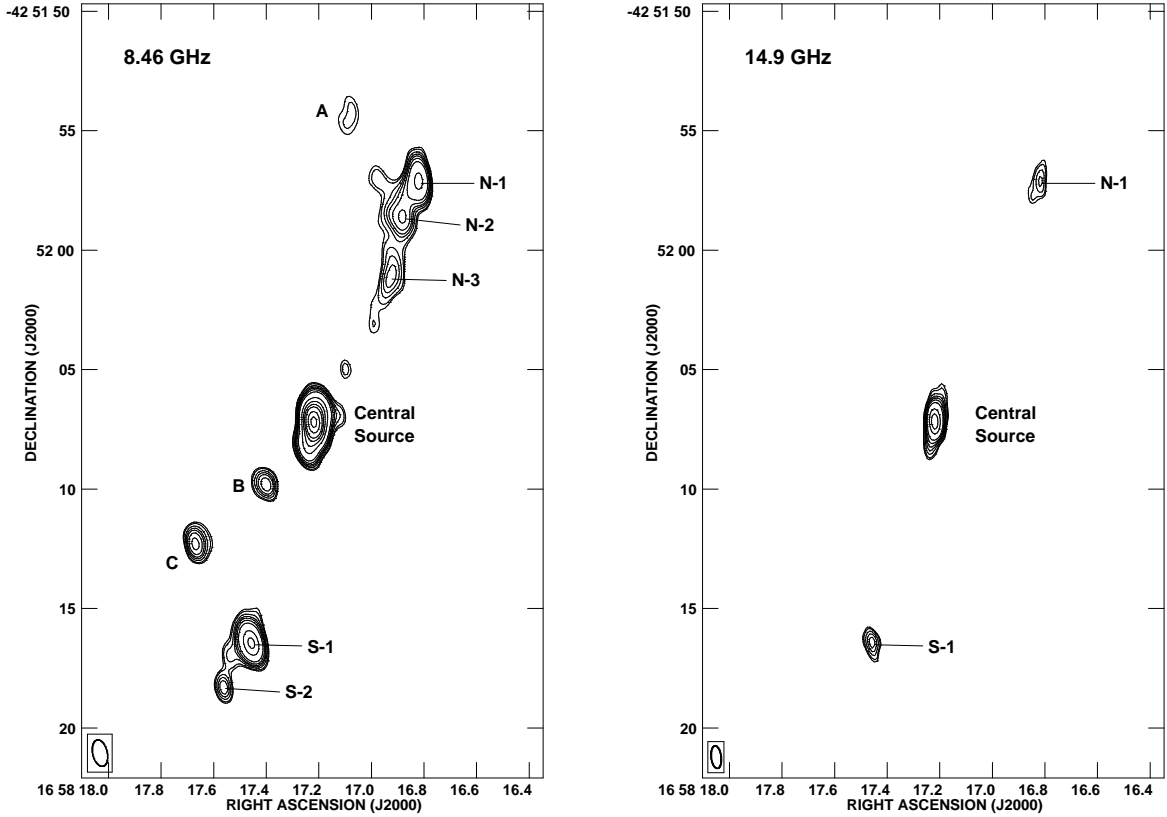}{f2b.eps}
\caption{VLA images at 8.46 (left) and 14.9 GHz (right) towards 
IRAS~16547-4247.
Contours are -5, 5, 6, 8, 10, 12, 15, 20, 40, 60,
80, 100, 140, and 180 times
30 and 87 $\mu$Jy beam$^{-1}$,
the rms of the 8.46 and 14.9 GHz images, respectively.
The half power contour of the synthesized beams
($1\rlap.{''}13 \times 0\rlap.{''}61$; PA = $13^\circ$ for the 8.46 GHz image and
$0\rlap.{''}95 \times 0\rlap.{''}42$; PA = $6^\circ$ for the 14.9 GHz image)
are shown in the bottom left corner of each panel.
The 8.46 GHz image was made with a ROBUST=0 weighting, while the 14.9 GHz 
image was made with a ROBUST=5 weighting.
\label{fig2}}
\end{figure}

\clearpage

\begin{deluxetable}{lcc}
\tablewidth{10.0cm}
\tablecaption{Bootstrapped Flux Densities of the Phase Calibrator (1626-298)}
\tablehead{
\colhead{}  & \colhead{Wavelength} 
& \colhead{Flux Density}  
\nl
\colhead{Epoch}  & \colhead{(cm)}
& \colhead{(Jy)}}
\startdata
2003 September 25 & 3.6 & 1.490 $\pm$ 0.002 \\ 
2003 September 30 & 3.6 & 1.527 $\pm$ 0.003 \\
2003 September 25 & 2.0 & 1.819 $\pm$ 0.005 \\
2003 September 30 & 2.0 & 1.856 $\pm$ 0.006 \\
\enddata
\end{deluxetable}

\clearpage

\begin{deluxetable}{l l l c c c }
\tabletypesize{\small}
\tablecolumns{6}
\tablewidth{0pc}
\tablecaption{Observed parameters of main sources \label{tab3}}
\tablehead{
\colhead{Comp.}                              &
\colhead{$\alpha$(2000)\tablenotemark{a}}                  &
\colhead{$\delta$(2000)\tablenotemark{a}}                  &
\colhead{Frequency}                     & 
\colhead{S$_\nu$\tablenotemark{b}}                         &
\colhead{Deconvolved Size\tablenotemark{c}}                           \\
\colhead{}                        &
\colhead{($16^h~58^m$)}           &
\colhead{($-42^\circ$)}           &
\colhead{(GHz)}                   & 
\colhead{(mJy)}                   & 
\colhead{}  \\ 
}
\startdata
\cutinhead{Central source}
  & 17$\rlap.^s$216$\pm$0.006 & 52$'$ 07$\rlap.{''}$64$\pm$0$\rlap.{''}$05 &
     4.80  & 7.3$\pm$0.4 & 
     2\rlap.{''}0$\pm$0\rlap.{''}2$\times$1\rlap.{''}1$\pm$0\rlap.{''}1; 
     24$^\circ \pm$10$^\circ$ \\
  & 17$\rlap.^s$210$\pm$0.003 & 52$'$ 07$\rlap.{''}$48$\pm$0$\rlap.{''}$02 &
     8.64 & 8.2$\pm$0.2 & 
     1\rlap.{''}39$\pm$0\rlap.{''}06$\times$0\rlap.{''}80$\pm$0\rlap.{''}04; 
     9$^\circ \pm$4$^\circ$ \\
  & 17$\rlap.^s$218$\pm$0.001 & 52$'$ 07$\rlap.{''}$22$\pm$0$\rlap.{''}$01 &
     8.46 & 8.7$\pm$0.1 &
     1\rlap.{''}02$\pm$0\rlap.{''}01$\times$0\rlap.{''}22$\pm$0\rlap.{''}02;
     163$^\circ \pm$1$^\circ$ \\ 
  & 17$\rlap.^s$219$\pm$0.001 & 52$'$ 07$\rlap.{''}$16$\pm$0$\rlap.{''}$01 & 
     14.9 & 10.5$\pm$0.3 &
     0\rlap.{''}74$\pm$0\rlap.{''}02$\times$0\rlap.{''}22$\pm$0\rlap.{''}03;
     167$^\circ \pm$2$^\circ$ \\
\cutinhead{North lobe}
N-1... & 16$\rlap.^s$815$\pm$0.009 & 51$'$ 57$\rlap.{''}$44$\pm$0$\rlap.{''}$07 &
     8.64 & $1.8\pm0.2$  & 
     1\rlap.{''}29$\pm$0\rlap.{''}18$\times$0\rlap.{''}53$\pm$0\rlap.{''}07;
     28$^\circ \pm$10$^\circ$ \\
    & 16$\rlap.^s$821$\pm$0.001 & 51$'$ 57$\rlap.{''}$08$\pm$0$\rlap.{''}$01 &
    8.46 & 2.0$\pm$0.1 &
    1\rlap.{''}30$\pm$0\rlap.{''}03$\times$0\rlap.{''}37$\pm$0\rlap.{''}05;
    159$^\circ \pm$2$^\circ$ \\
    & 16$\rlap.^s$824$\pm$0.005 & 51$'$ 57$\rlap.{''}$17$\pm$0$\rlap.{''}$06 &
    14.9 & 2.2$\pm$0.4 &
    1\rlap.{''}59$\pm$0\rlap.{''}11$\times$0\rlap.{''}41$\pm$0\rlap.{''}06;
    168$^\circ \pm$3$^\circ$ \\
N-2... & 16$\rlap.^s$883$\pm$0.01 & 51$'$ 58$\rlap.{''}$63$\pm$0$\rlap.{''}$1 &
    8.64 & $1.6\pm0.3$  & 
    2\rlap.{''}74$\pm$0\rlap.{''}25$\times$0\rlap.{''}60$\pm$0\rlap.{''}13;
    17$^\circ \pm$4$^\circ$ \\
    & 16$\rlap.^s$893$\pm$0.004 & 51$'$ 58$\rlap.{''}$39$\pm$0$\rlap.{''}$03 &
    8.46 & $1.5\pm0.2$  & 
    1\rlap.{''}81$\pm$0\rlap.{''}06$\times$0\rlap.{''}97$\pm$0\rlap.{''}03;
    180$^\circ \pm$3$^\circ$ \\
N-3... & 16$\rlap.^s$912$\pm$0.01 & 52$'$ 02$\rlap.{''}$13$\pm$0$\rlap.{''}$1 &
    8.64 & $0.4\pm0.1$\tablenotemark{d}  & ---  \\
    & 16$\rlap.^s$931$\pm$0.004 & 52$'$ 01$\rlap.{''}$16$\pm$0$\rlap.{''}$03 &
    8.46 & $1.0\pm0.1$  & 
    1\rlap.{''}76$\pm$0\rlap.{''}06$\times$0\rlap.{''}54$\pm$0\rlap.{''}05;
    165$^\circ \pm$2$^\circ$ \\
\cutinhead{South lobe}
S-1... & 17$\rlap.^s$462$\pm$0.007 & 52$'$ 16$\rlap.{''}$91$\pm$0$\rlap.{''}$06 &
    4.80 & $4.1\pm0.2$  & 
    1\rlap.{''}92$\pm$0\rlap.{''}22$\times$1\rlap.{''}19$\pm$0\rlap.{''}14;
    6$^\circ \pm$12$^\circ$ \\
    & 17$\rlap.^s$441$\pm$0.004 & 52$'$ 16$\rlap.{''}$83$\pm$0$\rlap.{''}$03 &
    8.64 & $2.8\pm0.1$  & 
    0\rlap.{''}91$\pm$0\rlap.{''}16$\times$0\rlap.{''}31$\pm$0\rlap.{''}32;
    175$^\circ \pm$5$^\circ$ \\
    & 17$\rlap.^s$459$\pm$0.002 & 52$'$ 16$\rlap.{''}$43$\pm$0$\rlap.{''}$01 &
    8.46 & 2.8$\pm$0.1 &
    0\rlap.{''}66$\pm$0\rlap.{''}04$\times$0\rlap.{''}46$\pm$0\rlap.{''}02;
    20$^\circ \pm$7$^\circ$ \\
    & 17$\rlap.^s$459$\pm$0.003 & 52$'$ 16$\rlap.{''}$45$\pm$0$\rlap.{''}$03 &
    14.9 & 2.0$\pm$0.1 &
    0\rlap.{''}66$\pm$0\rlap.{''}07$\times$0\rlap.{''}41$\pm$0\rlap.{''}04;
    15$^\circ \pm$11$^\circ$ \\
S-2... & 17$\rlap.^s$561$\pm$0.003 & 52$'$ 18$\rlap.{''}$26$\pm$0$\rlap.{''}$03 &
    8.46 & 0.42$\pm$0.05 &
    ---\tablenotemark{e} \\ 
\enddata

\tablenotetext{a}{Peak position. Right ascension ($\alpha$) given in hours, 
minutes, and seconds, and declination ($\delta$), given in degrees, arcmin, 
and arcsec. The errors given are formal statistical errors. The absolute 
positional error of the images is estimated to be 0$\rlap .{''}$2.}
\tablenotetext{b}{Total flux density.}
\tablenotetext{c}{Deconvolved dimensions of the source: FWHM major axis $\times$
FWHM minor axis; position angle of major axis.}
\tablenotetext{d}{Peak flux density (mJy beam$^{-1}$).}
\tablenotetext{e}{Unresolved}

\end{deluxetable}

\clearpage

\begin{deluxetable}{l c c c c } 
\tablecolumns{5} 
\tablewidth{0pc} 
\tablecaption{Parameters of the sources away from the outflow axis \label{tab3}}
\tablehead{
\colhead{Source}                              &                  
\colhead{$\alpha$(2000)\tablenotemark{a}}     &
\colhead{$\delta$(2000)\tablenotemark{a}}     &
\colhead{Frequency}                     &
\colhead{S$_\nu$\tablenotemark{b}}      \\
\colhead{}                           & 
\colhead{($16^h~58^m$)}               &
\colhead{($-42^\circ$)}               &
\colhead{(GHz)}                   &
\colhead{(mJy)}                   \\
}
\startdata

A....  & 17$\rlap.^s$162$\pm$0.027 & 51$'$ 56$\rlap.{''}$87$\pm$1$\rlap.{''}$04 & 
4.80 & 1.7$\pm$0.3 \\
   & 17$\rlap.^s$069$\pm$0.025 & 51$'$ 56$\rlap.{''}$11$\pm$0$\rlap.{''}$18 & 
8.64 & 1.2$\pm$0.4 \\
   & 17$\rlap.^s$084$\pm$0.005 & 51$'$ 54$\rlap.{''}$55$\pm$0$\rlap.{''}$13 & 
8.46  & 0.38$\pm$0.05 \\
B....  & 17$\rlap.^s$423$\pm$0.047 & 52$'$ 10$\rlap.{''}$38$\pm$0$\rlap.{''}$64 &
4.80  & 0.44$\pm$0.05 \\
   & 17$\rlap.^s$404$\pm$0.002 & 52$'$ 09$\rlap.{''}$79$\pm$0$\rlap.{''}$02 &
8.46  & 0.67$\pm$0.05 \\
   &  17$\rlap.^s$407$\pm$0.01 & 52$'$ 09$\rlap.{''}$68$\pm$0$\rlap.{''}$09 & 
14.9  & 0.70$\pm$0.24 \\
C....  & 17$\rlap.^s$652$\pm$0.023 & 52$'$ 12$\rlap.{''}$98$\pm$0$\rlap.{''}$20 &
4.80  & 1.08$\pm$0.05 \\
   & 17$\rlap.^s$623$\pm$0.010 & 52$'$ 13$\rlap.{''}$08$\pm$0$\rlap.{''}$11 &
8.64  & 0.74$\pm$0.06 \\
   & 17$\rlap.^s$665$\pm$0.001 & 52$'$ 12$\rlap.{''}$26$\pm$0$\rlap.{''}$02 &
8.46 & 0.90$\pm$0.06 \\
   & 17$\rlap.^s$641$\pm$0.008 & 52$'$ 12$\rlap.{''}$05$\pm$0$\rlap.{''}$06 &
14.9 & 0.64$\pm$0.20 \\

\enddata

\tablenotetext{a}{Peak position. Right ascension ($\alpha$) given in hours, 
minutes, and seconds, and declination ($\delta$), given in degrees, arcmin, 
and arcsec. The errors given are formal statistical errors. The absolute 
positional error of the images is estimated to be 0$\rlap.{''}$2.}
\tablenotetext{b}{Total flux density.} 

\end{deluxetable}


\end{document}